\documentstyle [12pt] {article}
\title {COSMOLOGICAL CONSTANT AS THE COEFFICIENT OF QUANTUM TUNNELING IN THE UNIVERSE EXTERIOR}

\author{Vladan Pankovi\'c$^{\ast,\sharp}$,
Miodrag Krmar$^\ast$, Simo Ciganovi\'c$^\sharp$\\
$^\ast$Department of Physics, Faculty of Sciences, 21000 Novi
Sad,\\ Trg Dositeja Obradovi\'ca 4. , Serbia, vdpan@neobee.net \\
$^\sharp$Gimnazija, 22320 Indjija, Trg Slobode 2a, Serbia \\}

\date {}

\begin{document}
\maketitle

\vspace {0.5cm} PACS number: 98.80.-k, 98.80.Qc \vspace {0.5cm}

\begin {abstract}
In this work we suggest a simple model of the cosmological
constant as the coefficient of the quantum tunneling of vacuum
fluctuations (with wave length larger than Planck length) at tiny,
boundary spherical shell of the universe (with thickness
equivalent to Planck length and radius equivalent to scale
factor). Roughly speaking, given fluctuations can, by quantum
tunneling (i.e. scattering with a potential barrier with highness
equivalent to Planck energy and width proportional to,
approximately, three hundred Planck length) leave universe and
arrive in its exterior, i.e. multi-universe (in sense of Linde
chaotic inflation theory universe can be considered as a
causally-luminally connected space domain while its exterior can
be considered as a space domain without causal-luminal connections
with universe). It is in full agreement with usual quantum
mechanics and quantum field theory as well as WMAP observational
data (especially fine tuning condition).
\end {abstract}

\vspace{0.5cm}

As it is well-known [1], [2] astronomical WMAP data prefer
standard cosmological model of the spatially flat universe with
cosmological constant $\Lambda \simeq 1.2 10^{-35}s^{-2}$. On the
other hand such value of the cosmological constant sharply
contradicts to standard prediction of the quantum field theories
that consider cosmological constant as the result of the quantum
vacuum fluctuations [3], [4]. Simply speaking mentioned
contradiction represents "the biggest blunder", i.e. error in the
entire history of sciences characterized by 123 magnitude order.
(More precisely, there is observationally obtained fine tuning
condition $\frac {L^{2}_{\Lambda}}{L^{2}_{P}}\simeq 10^{123}$
where $L_{\Lambda}$ represents the length corresponding to
$\Lambda$.)

In this work, generalizing our previous results [5] on formally
classically interpreted cosmological constant as a coefficient of
the negative surface tension, we shall suggest a simple model of
the cosmological constant. Precisely, we shall present
cosmological constant as the coefficient of the quantum tunneling
of vacuum fluctuations (with wave length larger than Planck
length) at tiny, boundary spherical shell of the universe (with
thickness equivalent to Planck length and radius equivalent to
scale factor). Roughly speaking, given fluctuations can, by
quantum tunneling (i.e. scattering with a potential barrier with
highness equivalent to Planck energy and width proportional to,
approximately, three hundred Planck length) leave universe and
arrive in its exterior, i.e. multi-universe. In sense of Linde
chaotic inflation theory [6], [7] universe can be considered as a
causally-luminally connected space domain while its exterior can
be considered as a space domain without causal-luminal connections
with universe. It is in full agreement with quantum mechanics and
quantum field theory as well as WMAP observational data
(especially fine tuning condition).

As it is well-known Friedmann equation for flat universe with
cosmological constant in post-radiation epoch has the following
form
\begin {equation}
   (\frac {\frac {d{\it a}}{dt}}{{\it a}})^{2} = \frac { G \rho 8 \pi }{3} + \frac {\Lambda}{3}
\end {equation}
where ${\it a}$ represents the scale factor, $G$ - Newtonian
gravitational constant, $\rho$ - universe matter mass density and
$\Lambda$ - cosmological constant. Given equation can be simply
transformed in
\begin {equation}
   \frac {1}{2}(\frac {d{\it a}}{dt})^{2} = G \rho \frac {4\pi}{3}\frac {{\it a}^{3}}{\it a} + \frac {\Lambda}{6}{\it a}^{2}
\end {equation}
i.e. in
\begin {equation}
   \frac {1}{2}(\frac {d{\it a}}{dt})^{2} - G \frac {M}{{\it a}} - \frac {\Lambda}{6}{\it a}^{2}= 0
\end {equation}
where
\begin {equation}
   M = \rho \frac {4\pi}{3}{\it a}^{3}= const
\end {equation}
represents the mass of the usual matter in the universe.

Equation (3), without term with cosmological constant $\frac
{\Lambda}{6}{\it a}^{2}$, admits formally, as it is well-known
[8], a classical interpretation. Namely, term $\frac {1}{2}(\frac
{d{\it a}}{dt})^{2}$ can be considered as the classical kinetic
energy of a probe particle with unit mass that is placed at
surface of the sphere with radius ${\it a}$ and that moves
radially with speed $\frac {d{\it a}}{dt}$. Also, term $- G \frac
{M}{{\it a}}$ can be considered as the classical potential energy
of the gravitational interaction between given probe particle and
given sphere with homogeneously distributed mass $M$ over whole
volume. We suggested [5] that term with cosmological constant
$\frac {\Lambda}{6}{\it a}^{2}$, in (3) can be, formally,
classically interpreted too. Namely, it has form analogous to the
work of a negative surface tension of a liquid.

In this way equation (3) can be formally classically considered as
the zero total energy of given probe particle with unit mass whose
kinetic energy increases for reason of increase of the absolute
value of work of a negative surface tension.

Now, we shall suggest a more detailed, quantum interpretation of
mentioned term with cosmological constant.

Consider, formally speaking, a space boundary of the universe in
form of a tiny spherical shell with radius ${\it a}$ and thickness
$L_{P}$ representing Planck length.

Consider, further, a small part of given boundary representing an
elementary cube with linear dimension $L_{P}$  and volume
$L^{3}_{P}$. Within given elementary cube, according to quantum
field theory, vacuum fluctuations of all types, all wave lengths
and all directions of propagation can occur, i.e. be created.

If fluctuation holds wave length smaller than Planck length then,
roughly speaking, it will practically certainly interact with
virtual or real quantum systems in elementary cube which can be
considered as fluctuation annihilation.

But, if fluctuation holds wave length larger than Planck length,
then, roughly speaking, it will practically certainly leave given
elementary cube and cannot annihilate by interaction with virtual
or real quantum systems in this elementary cube. However, it can
annihilate by interaction with real or virtual quantum systems in
other elementary cubes in boundary shell or in volume of limited
by given shell, except in case if given fluctuation propagates
progressively radially toward, formally speaking, exterior of
universe. Namely, in sense of Linde chaotic inflation theory [6],
[7] universe can be considered as a causally-luminally connected
space domain while its exterior, i.e. multi-universe, can be
considered as a space domain without causal-luminal connections
with universe.

If fluctuation with wave length larger than Plank length
propagates progressively radially toward exterior of the universe,
then, simplifiedly speaking, a scattering between fluctuation and
universe exterior will occur. This scattering we shall modeled
simply quantum mechanically. Namely, since, simplifiedly speaking,
fluctuation cannot arrive in exterior of the universe this
exterior can be presented as an infinitely high potential barrier.
But, since within quantum field theory there is no larger energy
than Planck energy, previous potential barrier can be changed by a
square-well potential barrier with highness equivalent to Planck
energy $E_{P}$ and width (length) $L$ whose value can be
determined latter. It implies that fluctuation with wave length
larger than Planck length and, correspondingly, momentum smaller
than Planck momentum $P_{P}=\frac {E_{P}}{c}= \frac
{\hbar}{L_{P}}$, can do a quantum tunnel effect through potential
barrier with transparency coefficient
\begin {equation}
    T = \exp [- \frac {P_{P}L}{\hbar}] = \exp [- \frac {L}{L_{P}}]
\end {equation}
where $c$ represents speed of light and $\hbar $ - reduced Planck
constant. In other words, there is a quantum tunneling of the
vacuum fluctuations with wave length larger than Plank length and
corresponding energy transition from universe in the universe
exterior.

Now, we shall simply demonstrate that mentioned quantum tunneling,
precisely transparency coefficient, can be used as a consistent
model of the cosmological constant. Namely, total energy that
universe loses by mass unit by described tunnel effect can be
roughly presented by expression
\begin {equation}
   E = (\frac {4\pi {\it a}^{2}L_{P}}{ L^{3}_{P}}) (\frac {1}{6}) T (\frac {mc^{2}}{m}) =
   (\frac {4 \pi {\it a}^{2}}{ L^{2}_{P}}) (\frac {c^{2}}{6}) \exp [-\frac {L}{L_{P}}]    \hspace{0.5cm}   {\rm for}  \hspace{0.5cm}   m \leq \frac {\hbar}{L_{P}c} .
\end {equation}
Here term $4\pi {\it a}^{2}L_{P}$ represents the boundary shell
volume, $\frac {4\pi {\it a}^{2}L_{P}}{ L^{3}_{P}}$ - number of
the elementary cubes in shell, $\frac {1}{6}$ probability that
vacuum fluctuation from elementary cube propagates progressively,
radially, toward universe exterior, $T = \exp [- \frac
{L}{L_{P}}]$ - transparency coefficient (5) and, $ mc^{2}/{m}$ -
energy of the fluctuation by mass unit in one elementary cube.

Expression (6) we can compare with term with cosmological constant
$\frac {\Lambda}{6}{\it a}^{2}$ in (3) which yields
\begin {equation}
   \Lambda= \frac {4\pi T c^{2}}{L^{2}_{P}} = \frac {c^{2}}{\frac {L^{2}_{P}}{4\pi T}} = \frac {c^{2}}{\frac {L^{2}_{P}}{4\pi \exp [-\frac {L}{L_{P}}]}} = \frac {c^{2}}{L^{2}_{\Lambda}}
\end {equation}
where
\begin {equation}
   L^{2}_{\Lambda} = \frac {L^{2}_{P}}{4\pi \exp[-\frac {L}{L_{P}}]}\simeq \frac { L^{2}_{P}}{\exp[-\frac {L}{L_{P}}]} .
\end {equation}

From (8) it follows
\begin {equation}
   \frac {L^{2}_{P}}{L^{2}_{\Lambda}} \simeq \exp [- \frac {L}{L_{P}}]             .
\end {equation}
In can be compared with observationally obtained fine tuning
condition
\begin {equation}
   \frac {L^{2}_{P}}{L^{2}_{\Lambda}} \simeq 10^{-123}             .
\end {equation}
Expression (9) and (10) can be equivalent for
\begin {equation}
    L \simeq 123 \ln [10] L_{P} = 123 \cdot 2.3 L_{P}\simeq 283.22 L_{P} \simeq 300 L_{P}                 .
\end {equation}
In this way we determined $L$ for which fine tuning condition is
satisfied and which represents a length characteristic for usual
quantum field theory.

In conclusion we can repeat and point out the following. In this
work we suggested a simple model of the cosmological constant as
the coefficient of the quantum tunneling of vacuum fluctuations
(with wave length larger than Planck length) at tiny, boundary
spherical shell of the universe (with thickness equivalent to
Planck length and radius equivalent to scale factor). Given
fluctuations can, by quantum tunneling (i.e. scattering with a
potential barrier with highness equivalent to Planck energy and
width proportional to, approximately, three hundred Planck length)
leave universe and arrive in its exterior, i.e. multi-universe (in
sense of Linde chaotic inflation theory). It is in full agreement
with usual quantum mechanics and quantum field theory as well as
WMAP observational data (especially fine tuning condition).
Nothing more is explicitly necessary for a simple and consistent
interpretation of the cosmological constant.

\vspace{1cm}

 {\Large \bf References}

\begin {itemize}

\item [[1]] D. N. Spergel et al., Astrophys. J. Supp. {\bf 146} (2003) 175 ; astro-ph/0302209.
\item [[2]] D. N. Spergel et al., Astrophys. J. Supp. {\bf 170} (2007) 337 ; astro-ph/0603449.
\item [[3]] S. Weinberg, Rev. Mod. Phys. {\bf 61} (1989) 1
\item [[4]] P. J. E. Peebles, Rev. Mod. Phys. {\bf 75} (2003) 559
\item [[5]] V. Pankovic, S. Ciganovic, J. Ivanovic, R. Glavatovic, P.Grujic, {\it A Simple Holographic Model of the Cosmological Constant}, astro-ph/0801.2859
\item [[6]] A. D. Linde, in, {\it 300 Years of Gravitation}, eds.  S. W. Hawking, W. Israel (Cambridge University Press, Cambridge, England, 1989.)
\item [[7]] A. D. Linde, {\it Inflation and Quantum Cosmology} (Academic Press, Boston, 1990.)
\item [[8]] B. W. Carroll, D. A. Ostlie, {\it An Introduction to Modern Galactic Astrophysics and Cosmology}, (Addison-Wesley, Reading, MA, 2007)

\end {itemize}

\end {document}